\documentclass[article]{JACoW-GSI}
\usepackage{graphicx}
\usepackage{psfrag}
\usepackage{url}

\begin{document}

\title{\hfill {\small FZJ-IKP-TH-2009-36, HISKP-TH-09-38} \\
Exotic open and hidden charm states \\ \vspace{-0.5cm}}
\author[1]{C. Hanhart\thanks{c.hanhart@fz-juelich.de}}
\author[1]{F.-K. Guo\thanks{f.k.guo@fz-juelich.de}}
\author[1,2]{U.-G. Mei\ss ner\thanks{meissner@hiskp.uni-bonn.de}}

\affil[1]{Forschungszentrum J\"ulich, Germany  }

\affil[2]{University of Bonn, Germany  }

\maketitle

\section{General remarks}

\begin{figure*}[tb]
\centering
\includegraphics*[width=145mm]{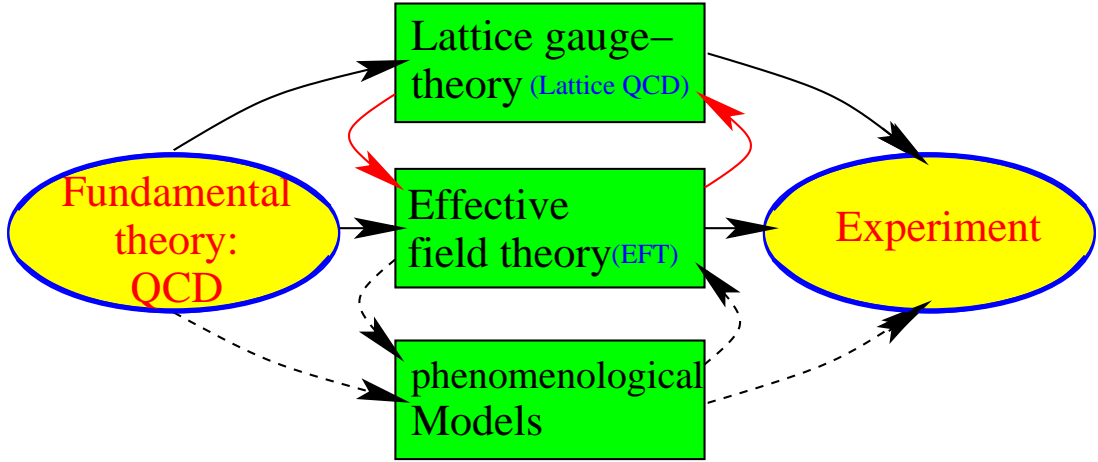}
\caption{Illustration of different methods to
compare properties of QCD to experiment.}
\label{fig:general}
\end{figure*}

To achieve a quantitative understanding of the theory of strong
interactions, QCD, at low energies, where the theory is
non--perturbative, is one of the toughest challenges of todays particle
physics. On the theoretical side in the last decades tremendous
progress was achieved by two developments, namely effective field
theories (EFTs) as well as lattice gauge theory. In the former the
symmetries of the fundamental theory are mapped on an effective theory
formulated in terms of physical degrees of freedom (here $\pi$, $K$,
$\eta$, $D$, $D^*$, $J/\psi$, ...) in a systematic and well-defined
way based on a power counting in pertinent small parameter(s). In the
latter the theory is solved on a discretized space--time. Both
approaches have a clear connection to QCD, but they also have their
drawbacks: Calculations within lattice QCD are numerically very
demanding and therefore a significant amount of expensive computing
time is necessary. For EFT calculations on the other hand need, when
pushed to higher and higher accuracy, an increasing number of
parameters --- the so--called low energy constants (LECs) --- needs to
be determined, either by direct experimental input or using dispersion
theory.\footnote{We note in passing that under certain circumstances EFT and 
dispersion relations can be combined successfully to achieve very precise
predictions at low energies.}
In recent years it became more and
more clear that the future of the field lies in the combination of
both these approaches, see e.g. \cite{Meissner:2006ai}.
On the one hand, the LECs of the
effective field theory related to the explicit symmetry breaking can
be determined from varying the quark masses on the lattice - provided
one is close enough to the chiral regime.  In this way from a few
lattice simulations in principle a large number of observables can be
deduced, since the LECs typically contribute to various
observables. Recent simulations allow to pin down the LECs related to
the quark mass dependence of the pion mass and the pion decay constant
to good precision, see e.g. \cite{Kadoh:2008sq} (and references therein). 
Here, one can use the chiral perturbation theory expressions for these 
quantities to extract the LECs from the measured quark mass dependence. 
Another example of this
interplay will be discussed below. 
On the other hand,
in general lattice calculations are
performed, at unphysically high quark masses, where computations are a
lot cheaper.  These can be connected to the physical parameter space
using extrapolation formulas derived using an appropriately 
tailored
EFT. This strategy is illustrated in Fig. \ref{fig:general}. 
 EFTs also allow one to systematically explore the volume and the
lattice spacing dependences that arise naturally in any simulation.
For a pedagogical introduction, see \cite{Sharpe:2006pu}.

In addition to the main lines from EFTs and lattice QCD, there
is also shown the path of phenomenological models. The
large number of existing quark models fall into this class.
The dashed arrows indicate that for those models there is
some connection to QCD --- often the phrase
``QCD inspired models'' is used ---  and observables can be extracted,
however, neither is the connection to QCD as rigorous as in
the previous two examples nor is it normally possible to
systematically improve the calculations. Still, phenomenological
models like the non--relativistic quark model,
can give important insights and are sometimes even used
to estimate the free parameters of the EFTs.

In addition to lattice QCD and EFTs, that allow one to calculate
observables from first principles, in special cases also general
theorems help to analyze data or to interpret results of
calculations. One example of such a theorem applies, if the resonance
of interest is located close to an $s$--wave threshold.  Then the
molecular admixture of this resonance by the corresponding continuum
channel can be quantified model-independently. The resulting analysis
method is outlined in the next section and applied to various states
below. It should be clear that any field theoretical approach has to
be consistent with such a general theorem. This holds especially for
EFTs as well as lattice QCD. In this case one may use the mentioned
theorem to interpret the results of some given calculation.  An
example of this is given below.

In recent years a large number of narrow states was found in the charm--sector
that do not fit into the conventional quark--antiquark picture.
An equally large number of non--conventional explanations was proposed for those,
including hybrids, glueballs, tetraquarks as well as hadronic molecules (for
reviews, see e.g.
Refs.~\cite{charmreview1,charmreview2,charmreview3,charmreview4,charmreview5}),
the last ones sometimes appearing as hadro--charmonia --- bound systems
of $c\bar c$ mesons and a light cloud~\cite{voloshin}.  At present
there are no model-independent methods available to disentangle the various
scenarios mentioned, besides for the hadronic molecules as outlined in the
previous paragraph. Thus, we will focus here on this kind of exotic states.

\section{Identifying molecular states}
\label{sec:ims}

Already 40 years ago Weinberg showed how the molecular content of
the deuteron can be quantified~\cite{wein}. In Ref.~\cite{evidence}
it was shown that this result can be generalized even to unstable
states as long as their width is narrow and the inelastic thresholds
are sufficiently far away. The central result of these studies is
that the effective coupling constant of a resonance to a continuum
channel in the $s$--wave can be written as
\begin{center}
\begin{eqnarray} \nonumber
{ \frac{g^2_{\rm eff}}{4\pi}}
 &=&
{4(m_1+m_2)^2{\lambda^2}\sqrt{2\epsilon/\mu}}
+{\cal O}\left(\frac{\sqrt{2\mu\epsilon}}{\beta}\right) \\
 &\le&
 4(m_1+m_2)^2\sqrt{2\epsilon/ \mu}
+{\cal O}\left(\frac{\sqrt{2\mu\epsilon}}{\beta}\right)
\label{weinform}
\end{eqnarray}
\end{center}
where $\epsilon=m_1+m_2-M_R$ denotes the binding energy of the
resonance $R$, measured relative to the continuum threshold of
particle $1$ and $2$, characterized here by their masses $m_1$ and
$m_2$. Further $\mu=m_1m_2/(m_1+m_2)$ is the reduced mass and
$\beta$ denotes the range of forces. Clearly, the formulas given are
useful only if the resonance pole is located {\sl very near a threshold},
for only then the model-dependent corrections that involve the range
of forces can safely be neglected. It should also be stressed that
for all those states, where the nearby continuum channel is not in
an $s$--wave, the scheme can not be applied.
The most important parameter in Eq.~(\ref{weinform})
is $\lambda^2$. It gives the probability to find the molecular
state in the physical state. Since the effective coupling
is an observable, as will also be illustrated in the next
paragraph, with Eq.~(\ref{weinform}) one is in the position
to ``measure'' the amount of the molecular admixture in some
physical state. Note, this implies especially that
the {\it structure information is hidden in the effective
coupling constant, independent of the phenomenology
used to introduce the pole(s).}
Therefore, a successful description of data using some model
does not necessarily mean that it is based on correct assumptions.
For example, if certain spectra can be described using
some isobar model --- a model where all dynamics is
parameterized by pole terms --- it does not mean that
all poles deduced are indeed genuine. The poles can
as well parameterize singularities of the $S$--matrix
generated by hadron--hadron dynamics. A nice illustration
of this fact is that one can very well study low
energy nuclear physics in an effective theory with
an explicit deuteron field~\cite{kaplan}. Also in
this case, the information on the {\it true nature}
of the deuteron is hidden in the effective coupling
constants in the sense discussed above. 

\section{The $Y(4660)$ and $Y_\eta(4616)$}

In order to illustrate how the method described in the previous
paragraph works, in this section we will apply it to the $Y(4660)$.
This state, discovered using ISR --- fixing the quantum numbers to
those of the photon, namely $J^{PC}=1^{--}$ --- in the $\psi'
\pi\pi$ final state ~\cite{Ydata}, is located very near the
$f_0(980)\psi'$ threshold. It is therefore natural to ask, if it is
an $f_0\psi'$ molecule.
An important feature of the $Y(4660)$ is that it was
neither observed in $e^+e^-\to\gamma_{ISR}\pi^+\pi^-J/\psi$
\cite{ISRJpsi}, nor in the exclusive $e^+e^-\to D{\bar D},D{\bar
D}^*, D^*{\bar D}^*, D{\bar D}\pi$ cross sections \cite{Exp:DDbar},
nor in the process $e^+ e^-\to J/\psi D^{(*)} {\bar D}^{(*)}$
\cite{Abe:2007sy}.  In the $f_0\psi'$ molecular picture, these facts
are easy to understand since the molecule would decay mainly through
the decays of the unstable $f_0(980)$. However, they would severely
challenge other models which try to explain the $Y(4660)$ as, for
instance, a canonical $5^3S_1$ $c{\bar c}$ charmonium~\cite{Ding} or a
baryonium~\cite{Qiao}.

\begin{figure}[htb]
\centering
\includegraphics*[width=65mm]{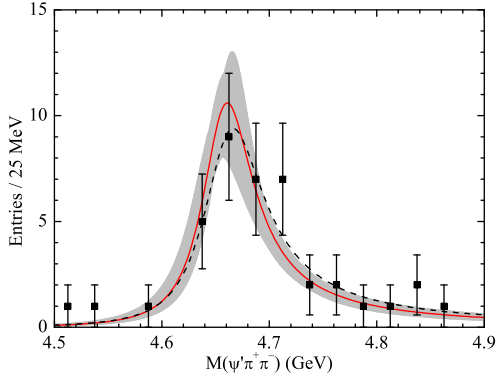}
\caption{Line shape for the $Y(4660)$ under the assumption of a
predominantly molecular nature. The best fit is shown as the solid
line, the grey band reflects the 1-$\sigma$ uncertainty of the
fit. The dashed line emerges as best fit, if also $g_{\rm eff}$ is
allowed to float.} \label{fig:Yspec}
\end{figure}

 As outlined above, for a given binding
energy and given masses for the constituents, the effective coupling
constant of a molecule can be calculated. Thus, it does no longer
appear as a parameter in the fit. For the analysis we developed an
improved Flatt\'{e} parameterization that allowed for a consistent
inclusion of the spectral function of the $f_0$ via a dispersion
integral --- for details see Ref.~\cite{GuoY}.  As parame\-ters in the
fit to the data now only the mass of the $Y$ and the overall
normalization appeared. The fit gave
\begin{equation}
M_Y=\left(4665^{+3}_{-5}\right) \ \mbox{MeV} \ .
\end{equation}
Clearly, by construction the mass of the $Y(4660)$ is forced
to stay below the $f_0\psi'$ threshold.
The result of the fit, shown as the grey band,
 is compared to the data
of Ref.~\cite{Ydata} in Fig.~\ref{fig:Yspec}.
The fit is very good. Even the slight asymmetry visible in
the data comes out naturally.
 To confirm that we were not misled by the fitting strategy
 we performed a second fit, where the effective coupling constant
 was allowed to float as well. Also this fit called for values
 of the couplings very near those extracted previously based on
 the molecular assumption, however, this more relaxed fit allowed
 for a larger range of masses for the $Y$.

In Ref.~\cite{GuoYeta} it was argued that, {\sl if} the $Y(4660)$ were
an $f_0\psi'$ molecule, heavy quark spin symmetry predicts a
pseudoscalar partner at $M_{Y_\eta}=M_Y-(M_{\psi'}-M_{\eta_c'}) $
being an $f_0\eta_c'$ molecule with very similar properties.
Especially the line shape in the $\eta_c'\pi\pi$ channel should be nearly 
identical to that shown in
Fig.~\ref{fig:Yspec}. The mass prediction, $M_{Y_\eta}= 4616^{+5}_{-6}\,$MeV, 
should hold with very high
accuracy, since the spin symmetry breaking terms in case of hadronic
molecules with a $c\bar c$ constituent are suppressed by an
additional factor $1/m_c^2$ --- this emerges since those
interactions need to be spin-dependent, connect colour neutral
objects, and quark exchange is not possible. The width of the
$Y_\eta$ into $\eta_c' \pi\pi$ is predicted to be 
$\Gamma(\eta_c'\pi\pi)=(60\pm30)\,$MeV.
An experimental
confirmation of this prediction would be very desirable for it would
provide further evidence for the picture sketched here.

\section{$D_{s0}^*(2317)$}
\label{sec:Ds}

So far  a method was described that allows one, under certain
conditions, to spot  molecules under the many states observed.
However, the whole scheme was based on non--relativistic quantum
mechanics. What we learn from the above is only additional
information which states QCD forms, but no direct connection to QCD
was employed. As outlined in the introduction, this is possible
systematically only using either lattice QCD or effective field
theories --- more precisely chiral perturbation theory to control
the dynamics of mesons formed of light quarks as well as heavy quark
effective field theory to construct the interactions amongst heavy
fields and, of course, suitable combinations thereof for heavy-light 
systems.

In this section we will study resonances formed of a light and a
heavy meson. However, no perturbative treatment is able to produce
poles. Thus we need to leave the ground of rigorous effective field
theories and add on top a resummation scheme. The power counting is done
of the level of the effective potential so that the convergence 
of the series has to be checked carefully for the resulting scattering
amplitudes and bound state properties, much in the sense outlined by
Lepage more than a decade ago \cite{Lepage}. This resummation scheme at the same time
has the beneficial side effect that all scattering amplitudes are
unitary. Note, this procedure is completely analogous to what is
routinely done in case of few--nucleon scattering --- for a recent
review see Ref.~\cite{fewNref}.

More precisely, we focus on the scattering of
$\pi$, $K$, and $\eta$ off $D$--mesons --- see Refs.~\cite{GuoD2,GuoD1}. Our
method is very similar to that of Ref.~\cite{LS},
however, for the first time higher order operators
are studied completely and a comparison to lattice data
is performed. In addition the poles in the complex plane
were identified revealing similar patterns in the
heavy-light systems as observed previously for
light resonances in Ref.~\cite{withjose}.

\psfrag{Vorhersage}{{}}
\psfrag{mp}{$m_\pi$}
\psfrag{c1}{$D\pi \ (I=3/2)$}
\psfrag{c2}{$D\bar K \  (I=1)$}
\psfrag{c3}{$D_s K \ (I=1/2)$}
\psfrag{c4}{$D K \ (I=0)$}
\begin{figure*}[tb]
\centering
\includegraphics*[width=145mm]{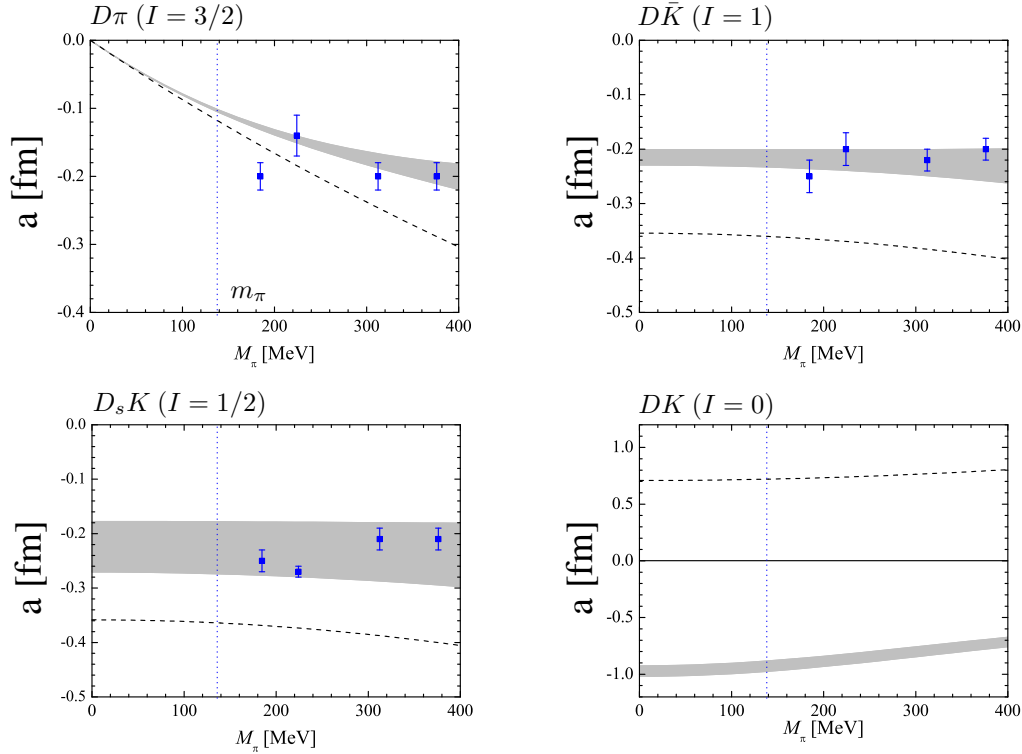}
\caption{Goldstone boson scattering off $D,D_s$-mesons in
unitarized chiral perturbation theory. 
The dashed line shows the result of leading order chiral
perturbation theory, while the grey band denotes the results of the
full calculation. The lattice data are from Ref.~\cite{lattice}.}
\label{fig:extrapol} 
\end{figure*}

It is especially important that a complete next--to--leading order
calculation was performed which allowed for the first time
for a reliable uncertainty estimate, which is, as we will see,
especially important in case of the strong width of the
$D_{s0}^*(2317)$, for here the quark model prediction is
significantly lower~\cite{godfrey,fulvia} than that derived within
the molecular picture. The scalar charmed--strange meson
$D_{s0}^*(2317)$ is of particular interest since its observed mass
is much lower than the expected value from the non-relativistic
quark model~\cite{Godfrey:1985xj}. Some ideas were proposed to shift
the scalar charmed-strange meson mass down to the mass range of the
$D_{s0}^*(2317)$. In~\cite{Lakhina:2006fy}, one loop corrections to
the spin-dependent one-gluon exchange potential were considered.
Another kind of modification is the mixing of the $c{\bar s}$ with
the $cq{\bar s}{\bar q}$ tetraquark~\cite{Browder:2003fk}, or
considering the coupling of the $c{\bar s}$ to hadronic channels,
such as $DK$~\cite{van Beveren:2003kd}. However, such kind of mass
shift calculations are model-dependent as pointed out in
Refs.~\cite{Capstick:2007tv,Guo:2007up}.  A $KD$ molecular
interpretation was proposed by Barnes et al.~\cite{Barnes:2003dj},
which agrees with a lot of dynamical calculations~\cite{Hchua,Guo1}.
Other exotic explanations were also proposed, such as tetraquark
state~\cite{tetraq}, and $D\pi$ atom~\cite{Dspi}. Thus, in order to
decide if an experiment can be decisive on the nature of this state,
it is necessary that some measure is found to quantify the
reliability of the results.

Since the isoscalar $D_{s0}^*(2317)$ is located below the $DK$
threshold, the only open strong decay channel is $D_s\pi^0$, which
calls for isospin violation. In QCD there are two sources of isospin
violation, namely quark mass differences and electromagnetism.
Typically both contribute with similar strength, thus a scheme is
necessary for the decays of the $D_{s0}^*(2317)$ that allows for a
consistent inclusion of both type of effects. This framework is given by
chiral perturbation theory. We performed a calculation up to next-to-leading
order (NLO). It
turned out, although essential for a quantitative understanding of
the $D^0$, $D^+$ mass difference, electromagnetic effects played a
minor role in the strong decay of the $D_{s0}^*(2317)$ --- thus the
decay was dominated by $\pi^0-\eta$ mixing, first shown in
Ref.~\cite{Guo1} to give a width of 8~keV, as well as the $D$--meson
mass differences in the loops, studied previously in
Refs.~\cite{tuebingen,LS}, which, when added in, give a width of 180~keV. 
On the contrary in case of a pure quark state the meson
loop contribution is absent and thus a width of about 8~keV is
predicted~\cite{fulvia}. Similar values are expected for 
all other non--molecular scenarios.
Thus, there is a direct experimental test available
for the conjecture that the  $D_{s0}^*(2317)$
is a hadronic molecule. 

Unfortunately, the uncertainty of 110~keV turns out to be quite
large. However, this uncertainty originates solely form the poorly
determined isospin conserving interactions --- to the order we are
working all isospin breaking terms could be fixed from other
sources. However, there are other tools to reduce the uncertainty on
the isospin conserving part, namely by comparing to lattice QCD.  For
this comparison scattering lengths (i.e the scattering amplitudes at
threshold) are ideal: on the one hand they can be calculated on the
lattice by systematic studies of the volume dependence of
correlators~\cite{luescher} and on the other hand they can be
straightforwardly calculated on the basis of (unitarized) chiral
perturbation theory. The predictions of our calculation are shown in
Fig.~\ref{fig:extrapol} as a function of the pion mass. The dashed
line shows the result of leading order chiral perturbation theory,
while the grey band denotes the results of the full calculation. The
lattice data are from Ref.~\cite{lattice}. Note that in the lattice
calculations only the light quark mass was varied while the strange
quark mass was kept fixed close to its physical value --- this is why
for vanishing $m_\pi$ only the scattering length for $\pi D$
scattering vanishes, as demanded by the Goldstone theorem. As can be
seen, in all channels our calculation agrees very well with the
lattice results. The expansion appears to converge very well for $\pi
D$ scattering, while in case of $D\bar K$ and $D_s K$ scattering in
the $I=1$ and $I=1/2$ channels, respectively, up to 50\% corrections
emerged from higher orders. Although quite sizable, even this change
is within the expectations from the power counting as a result of the
relatively large kaon mass. However, in case of $KD$ scattering in the
isoscalar channel the change is truely dramatic --- the scattering
length even changes its sign. This change in sign, clearly a
non--perturbative effect, is a consequence of the appearance of the
$D_{s0}^*(2317)$ as a $KD$ bound state.  In this context it is
important to remember the theorem stated above that for bound states
the effective coupling constant can be calculated directly from
masses, the binding energy and $\lambda^2$, the probability to find
the molecular state in the physical state ---
c.f. Eq.~(\ref{weinform}). This property translates directly into a
prediction for the scattering length, if the binding energy of the
bound state is small.  In this case the scattering length may be
written as
\begin{equation}
a \simeq
 \frac{-2\lambda^2/(1+\lambda^2)}{\sqrt{2m_K\epsilon}} \ .
\label{alam}
\end{equation}
And indeed, the value we predict for the
$KD$ scattering length exactly agrees with the prediction from the
general theorem in case of a purely molecular nature of the
$D_{s0}^*(2317)$ ($\lambda^2=0$). This opens the opportunity to
extract the nature of the $D_{s0}^*(2317)$ directly from a lattice
study: if this study were to find a scattering length in this
channel of about $-1$~fm, it would prove the nature of the
$D_{s0}^*(2317)$ as a $KD$ molecule. A larger absolute value of the
scattering length is not allowed, while any smaller value would
immediately quantify a non--molecular admixture
--- see Eq.~(\ref{alam}). This equation also embodies
the concept of universality for large scattering length,
for a lucid review see \cite{Braaten:2004rn}.

Above we argued that lattice data can
be used to determine the LECs of chiral perturbation
theory. How this could work in the future can also be
read off Fig.~\ref{fig:extrapol}. At present both
the theoretical predictions as well as the lattice data
show a significant scatter.
In addition, so far the
lattice studies were performed only for one lattice 
spacing. Further investigations are necessary to
quantify the systematics of the analysis and also adding 
more points for pion masses below 300~MeV. However, 
we hope that soon the lattice data/analysis will  improve
and then the parameters of our theoretical study can
be constrained by a fit to those data (this calls 
for a complete theoretical calculation to NNLO, however,
which still needs to be done). Once this is done,
one can expect that the theoretical uncertainty of
the prediction for the width of the $D_s(2317)$ is
reduced significantly as well.

\section{Discussion}

If mesonic bound states exist at all, shouldn't one expect a large
number of those in the spectrum? First of all one should stress that
hadron dynamics normally produces sufficient attraction only in
very few channels. This can also be seen, e.g., in
Fig.~\ref{fig:extrapol}: in all channels displayed, besides the $KD$
$I=0$ channel, where the $D_{s0}^*(2317)$ can be found, the
interaction is repulsive. In addition, there may already exist various
molecules in both, the light sector (here, e.g., the $f_0(980)$ is a
prominent candidate for a $\bar KK$ molecule~\cite{IaW}, not only
because of its mass close to the $\bar KK$ threshold but also because
of its properties~\cite{evidence,ourf0}) as well as in the charm
sector. Here the most prominent candidate not discussed in this
contribution is the $X(3872)$ as a prime candidate for a $D^*\bar
D+h.c.$ molecule (see Ref.~\cite{braaten} and references therein) or
virtual state, possibly with some $c\bar c$ admixture (see
Refs.~\cite{ourX} and references therein). 
In this contribution we outlined under what circumstances it is possible
to get model-independent statements about the molecular nature of
states --- this is especially possible for states located very close
to an $s$--wave threshold. However, models predict many more
states. E.g. a brother of the $Y(4660)$ discussed above as a candidate
for a $f_0\psi'$ bound state, could be the $Y(4260)$ in
Ref.~\cite{osetnew} claimed to the a resonance of $f_0(980)$ and
$J/\psi$.  In Refs.~\cite{Wang:2009hi,Wang:2009cw} similar conclusions
are drawn based on QCD sum rules.

On the long run we will be able to address these issues, namely
through the mentioned interplay of general theorems, lattice QCD as
well as effective field theories. As was shown in
Ref.~\cite{withjose}, there are cases where it is possible to move a
resonance, by changing a QCD parameter --- here the quark mass ---
from a position deep in the complex plane to a kinematic situation
where the bound state analysis sketched above can be applied. The
precondition for this is clearly to have a formalism which allows
for a controlled quark mass dependence.\footnote{As a word of caution
we should add that a consistent power counting for resonances still
needs to be developed, see e.g.~\cite{Bruns}   and references therein.}
 The parametric dependence of
resonances on QCD parameters can also be checked in a study of the
corresponding form factors~\cite{withfelipe}. This investigation
provides an additional check of the systematics of the approaches.
In this sense we can hope to get a deeper understanding of QCD in
the non--perturbative regime, once improved experimental data are
available to check and refine the theoretical approaches.


\end{document}